\documentclass[amsmath,amssymb,floatfix,lengthcheck,preprintnumbers,prd,showpacs,superscriptaddress,twocolumn]{revtex4}
\usepackage{graphicx}
\usepackage{hyperref}
\usepackage{cleveref}
\usepackage{pstricks}

\newcommand{\beq}{\begin{equation}}
\newcommand{\eeq}{\end{equation}}

\newcommand{\Qssub}{\ensuremath{\mbox{$Q_{s,\textrm{sub}}$}}}
\newcommand{\errpm}[2]{$\left(\begin{smallmatrix}#1\\#2\end{smallmatrix}\right)$}

\begin{document}

\title{Maximum-Likelihood Approach to Topological Charge Fluctuations \\ in Lattice Gauge Theory}

%\author{T.~Appelquist}
%\affiliation{Department of Physics, Sloane Laboratory, Yale University,
%             New Haven, Connecticut 06520, USA}
%\author{E.~Berkowitz}
%\affiliation{Lawrence Livermore National Laboratory, Livermore, California 94550, USA}
\author{R.~C.~Brower}
\affiliation{Department of Physics, Boston University,
	Boston, Massachusetts 02215, USA}
\affiliation{Aspen Center for Physics, Aspen, CO 81611, USA}
%\author{M.~I.~Buchoff}
%\affiliation{Institute for Nuclear Theory, Box 351550, Seattle, WA 98195-1550, USA}
\author{M.~Cheng}
\affiliation{Center for Computational Science, Boston University,
	Boston, Massachusetts 02215, USA}
\affiliation{Aspen Center for Physics, Aspen, CO 81611, USA}	
\author{G.~T.~Fleming}
\affiliation{Department of Physics, Sloane Laboratory, Yale University,
             New Haven, Connecticut 06520, USA}
\affiliation{Aspen Center for Physics, Aspen, CO 81611, USA}
%\author{J.~Kiskis}
%\affiliation{Department of Physics, University of California,
%	Davis, California 95616, USA}
\author{M.~F.~Lin}
\affiliation{Argonne Leadership Computing Facility, Argonne National Laboratory, Argonne, IL 60439, USA}
\affiliation{Computational Science Center, Brookhaven National Laboratory, Upton,
NY 11973, USA}
\affiliation{Aspen Center for Physics, Aspen, CO 81611, USA}
\author{E.~T.~Neil}
\affiliation{Department of Physics,
        University of Colorado, Boulder, CO 80309, USA}
\affiliation{RIKEN-BNL Research Center, Brookhaven National Laboratory, Upton, NY 11973, USA}
\affiliation{Aspen Center for Physics, Aspen, CO 81611, USA}
\author{J.~C.~Osborn}
\affiliation{Argonne Leadership Computing Facility, Argonne National Laboratory, Argonne, IL 60439, USA}
\author{C.~Rebbi}
\affiliation{Department of Physics, Boston University,
	Boston, Massachusetts 02215, USA}
\author{E.~Rinaldi}
\affiliation{Lawrence Livermore National Laboratory, Livermore, California 94550, USA}
\author{D.~Schaich}
\affiliation{Department of Physics, University of Colorado, Boulder, CO 80309, USA}
\affiliation{Department of Physics, Syracuse University, Syracuse, NY 13244, USA}
\affiliation{Aspen Center for Physics, Aspen, CO 81611, USA}
\author{C.~Schroeder}
\affiliation{Lawrence Livermore National Laboratory, Livermore, California 94550, USA}
%\author{S.~Syritsyn}
%\affiliation{RIKEN-BNL Research Center, Brookhaven National Laboratory, Upton, NY 11973, USA}
\author{G.~Voronov}
\affiliation{Department of Physics, Sloane Laboratory, Yale University,
             New Haven, Connecticut 06520, USA}
\author{P.~Vranas}
\affiliation{Lawrence Livermore National Laboratory, Livermore, California 94550, USA}
\author{E.~Weinberg}
\affiliation{Department of Physics, Boston University,
	Boston, Massachusetts 02215, USA}
\author{O.~Witzel}
\affiliation{Center for Computational Science, Boston University,
	Boston, Massachusetts 02215, USA}
%\collaboration{Lattice Strong Dynamics (LSD) Collaboration}
%\noaffiliation

%%%%%%%%%%%%%%%%%%%%%%%%%%%%%%%%%%%%%%%%%%%%%%%%%%%%%%%%%%%%%%%%%%%%%%%%%%%%%%
\begin{abstract}
%%%%%%%%%%%%%%%%%%%%%%%%%%%%%%%%%%%%%%%%%%%%%%%%%%%%%%%%%%%%%%%%%%%%%%%%%%%%%%
We present a novel technique for the determination of the topological susceptibility (related to the variance of the distribution of global topological charge) from lattice gauge theory simulations, based on maximum-likelihood analysis of the Markov-chain Monte Carlo time series.  This technique is expected to be particularly useful in situations where relatively few tunneling events are observed.  Restriction to a lattice subvolume on which topological charge is not quantized is explored, and may lead to further improvement when the global topology is poorly sampled.  We test our proposed method on a set of lattice data, and compare it to traditional methods.
%%%%%%%%%%%%%%%%%%%%%%%%%%%%%%%%%%%%%%%%%%%%%%%%%%%%%%%%%%%%%%%%%%%%%%%%%%%%%%
\end{abstract}
%%%%%%%%%%%%%%%%%%%%%%%%%%%%%%%%%%%%%%%%%%%%%%%%%%%%%%%%%%%%%%%%%%%%%%%%%%%%%%


\preprint{LLNL-JRNL-650193}

\maketitle

%%%%%%%%%%%%%%%%%%%%%%%%%%%%%%%%%%%%%%%%%%%%%%%%%%%%%%%%%%%%%%%%%%%%%%%%%%%%%%
\section{Introduction \label{sec:intro}}
%%%%%%%%%%%%%%%%%%%%%%%%%%%%%%%%%%%%%%%%%%%%%%%%%%%%%%%%%%%%%%%%%%%%%%%%%%%%%%

Lattice field theory is a powerful technique for the numerical study of
Yang-Mills gauge theories.  Recovery of continuum field-theory results
requires extrapolations in the lattice spacing and volume, which are generally
controlled and well-understood.  One effect of working in a finite volume $V$
is that the theory becomes dependent on the global topological charge $Q$
\cite{Brower:2003yx, Aoki:2007ka}.  Locality and cluster decomposition
properties suggest that such effects vanish as $V \rightarrow \infty$, but
they must be accounted for in the extrapolation.

In Euclidean Yang-Mills quantum field theory on a torus, the topological
charge $Q$ is quantized, dividing the configuration space into distinct
topological sectors.  These sectors are separated by an action barrier, so that the use of
sampling algorithms which favor small changes in the action (such as the
commonly used hybrid Monte Carlo algorithm) can lead to poor sampling of this
distribution.  Since the action barrier can grow with decreasing lattice
spacing \cite{Alles:1996vn, DelDebbio:2004xh, Luscher:2010iy} or increasing
number of flavors $N_f$ \cite{Appelquist:2009ka, Appelquist:2012nz}, the cost of tunneling to
different topological sectors can vary greatly depending on the details of the
calculation.

The ``freezing'' of topological charge resulting from these algorithmic problems leads to extremely long autocorrelation times, so that the distribution of $Q$ is poorly sampled.  Correction of the resulting systematic effects on observables can be done \cite{Brower:2003yx, Aoki:2007ka}, but these corrections require as inputs the cumulants of the topological charge distribution, particularly the variance $\langle Q^2 \rangle \equiv V \chi_t$, where $\chi_t$ is the topological susceptibility.  Using the standard estimator for variance requires many independent samples;
autocorrelations can lead to relatively few independent measurements
and a biased estimate with a large sampling error.

In this work, we suggest two ways to proceed when confronted with this
problem.  First, it is generally believed that the maximum likelihood (ML)
method (see Sec.~36.1.2 of \cite{Beringer:1900zz}) can produce reliable
estimates of model parameters when there are relatively few independent
samples of a distribution, provided the functional form of the underlying
distribution is known analytically.  Inspired by the work of Phil Nelson and
collaborators \cite{Beausang:2011}, we present such a maximum likelihood approach
(following an example by Franco \cite{Franco:2003} in the context of financial
time-series) to analyze the complete time-series information $\{Q_n\}$.  The analysis is done without blocking, since the effect of autocorrelations is built into the model.  This
method in principle allows the estimation of $\chi_t$ from even a handful of tunneling
events.

Eventually, if one performs a calculation at sufficiently small (but finite)
lattice spacing \cite{Luscher:1981zq}, global topological charge will never
change in any finite number of Markov steps.  If we choose a lattice volume
$V$ such that $V \chi_t \gg 1$, we can consider the distribution of
topological charge $Q_s$ computed only in subvolumes $V_s \gg \chi_t^{-1}$,
which by locality and cluster decomposition should also be distributed asymptotically as the stationary distribution $P(Q)$.  In this scenario, we can employ
either our ML method or a more standard blocked sample variance estimate to
compute the susceptibility, depending on the number of independent samples \footnote{When this work was essentially complete, we were informed that this approach had been explored briefly in the past \cite{deForcrand:1998ng}.}. Empirically, we find that the calculation of $\chi_t$ based on a subvolume gives the most robust estimates of $\chi_t$ in the case that relatively few uncorrelated measurements of $Q$ are available, although further study of this approach is needed.

%We conclude with the observation that these two approaches are complementary because scenarios can arise where both methods may not be applicable.  \draftnote{(DS: clarify?  State conclusion that subvolume-OU seems best?)} For example, one might desire to perform a calculation at as small a lattice spacing as possible to suppress finite lattice spacing effects on some observable.  This will likely mean that $V \chi_t \sim 1$ so employing the subvolume approach seems questionable.  Yet a small lattice spacing will mean long autocorrelation times in the topological charge, suggesting the use of a maximum likelihood method.

The contents of this manuscript are as follows: in \cref{sec:distQ} we discuss what is known about the distribution of topological charge in lattice simulations of Yang-Mills gauge theories.  \Cref{sec:OU} gives the definition of an Ornstein-Uhlenbeck (OU) process, which uniquely describes continuous Markov processes that remain Gaussian distributed.  \Cref{sec:ML} describes the maximum-likelihood estimation of $\chi_t$ based on the OU model.  In \cref{sec:fixedtop}, the implications of studying calculations with nearly-fixed global topological charge are discussed, and a modification of the maximum-likelihood estimate using lattice subvolumes is introduced.  \Cref{sec:examples} demonstrates the use of the proposed methods to extract $\chi_t$ on an example set of lattice configurations, and compares to other standard approaches.  Finally, \cref{sec:discussion} summarizes our results and discusses future applications and possible improvements.

%%%%%%%%%%%%%%%%%%%%%%%%%%%%%%%%%%%%%%%%%%%%%%%%%%%%%%%%%%%%%%%%%%%%%%%%%%%%%%
\section{Distribution of topological charge \label{sec:distQ}}
%%%%%%%%%%%%%%%%%%%%%%%%%%%%%%%%%%%%%%%%%%%%%%%%%%%%%%%%%%%%%%%%%%%%%%%%%%%%%%

Numerical lattice computations make use of a Markov process to sample the configuration space, generating a sequence of configurations $U_0 \to U_1 \to
\cdots \to U_n$ with corresponding topological charges $\{Q_n\}$.  As the sample size $n$ increases, the distribution $P(Q_n)$ converges to a stationary distribution $P(Q)$.

What is known about the distribution $P(Q)$?  With zero $\theta$-parameter, all odd cumulants of the distribution must vanish by parity invariance.  Furthermore, analysis of SU$(N_c)$ gauge theories at large-$N_c$ shows that the even cumulants scale as $\kappa_{2n} \sim N_c^{2-2n}$ \cite{'tHooft:1973jz, Witten:1978bc, Witten:1980sp}, suggesting that the distribution will be approximately Gaussian.  Given the suppression of higher cumulants, it seems reasonable to express the distribution $P(Q)$ in terms of its Edgeworth series \cite{Blinnikov:1997jq}, truncated to the first non-Gaussian term:
\beq\label{eq:EdgeworthQ}
P(Q) = \frac{32}{32+\epsilon} P_G\left(\frac{Q}{\sqrt{\sigma}}\right) \left[
1 + \frac{\epsilon}{4!} \mathrm{He}_4\left(\frac{Q}{\sqrt{\sigma}}\right)
\right],
\eeq
where $P_G(x)$ is the Gaussian distribution with zero mean and unit variance
and $\mathrm{He}_4(x) = x^4 - 6 x^2 + 3$ is a Hermite polynomial.  The
variance $\kappa_2$ and 4th-order cumulant $\kappa_4$ for this distribution
are
\beq\label{eq:EdgeworthQstats}
\kappa_2 = \sigma^2 \left( 1 + \frac{\epsilon}{4} \right) + O(\epsilon^2),
\quad
\frac{\kappa_4}{\kappa_2^2} = \epsilon + O(\epsilon^2).
\eeq
We identify the variance $\langle Q^2 \rangle = \kappa_2 \equiv V \chi_t$, which defines the topological susceptibility $\chi_t$.  As $\epsilon \rightarrow 0$, this distribution becomes purely Gaussian; several lattice studies have empirically found non-zero $\epsilon$ in SU$(N_c)$ gauge theories \cite{D'Elia:2003gr, DelDebbio:2006df, Durr:2006ky, Giusti:2007tu, Bonati:2013tt}.  We note that the dependence of this non-Gaussianity on the presence of light fermions is unclear, and large-$N_c$ arguments may be inapplicable for theories with many fermions $N_f$, unless $N_f/N_c$ is held fixed as $N_c \rightarrow \infty.$

%%%%%%%%%%%%%%%%%%%%%%%%%%%%%%%%%%%%%%%%%%%%%%%%%%%%%%%%%%%%%%%%%%%%%%%%%%%%%%
\section{Ornstein-Uhlenbeck process \label{sec:OU}}
%%%%%%%%%%%%%%%%%%%%%%%%%%%%%%%%%%%%%%%%%%%%%%%%%%%%%%%%%%%%%%%%%%%%%%%%%%%%%%

We wish to consider Markov processes that can reproduce the approximately-Gaussian topological charge distribution \cref{eq:EdgeworthQ}.  In fact, up to linear transformations in the variables, there is a unique non-trivial example of a continuous Markov process in which the expected distribution at any point in the stochastic evolution is Gaussian: the Ornstein-Uhlenbeck (OU) process \cite{Uhlenbeck:1930zz, Doob:1942zz}, which describes the Brownian
motion of a massive particle in the presence of arbitrary linear friction.
This process is described by the stochastic differential equation
\begin{equation} \label{eq:OUstochastic}
\frac{d}{dt} x(t) = - \eta \left(x(t) - \overline{x}\right)
+ \sigma \frac{d}{dt} W(t)
\end{equation}
where $\eta > 0$, $\sigma > 0$ and $W(t)$ is the stochastic Wiener process of
Brownian motion.  The standard solution leads to the following statistics:
\begin{eqnarray}
\mathrm{E}[x(t)] & = & \overline{x}
+ \left( x(0) - \overline{x} \right) e^{-\eta t} \nonumber \\*
\mathrm{Var}[x(t)] & = & \frac{\sigma^2}{2\eta}
\left( 1 - e^{-2 \eta t} \right) \label{eq:OUvarmean}
\end{eqnarray}
which converge to a Gaussian with mean $\bar{x}$ and variance $\sigma^2 / 2
\eta$ as $t \rightarrow \infty$, independent of the starting position $x(0)$.  We will discuss the accuracy of this model in the presence of small non-Gaussianities in \cref{sec:ML} below.

The detailed evolution of topological charge in a lattice gauge theory
calculation is quite complex and dependent on unphysical details such as the
choice of the update algorithm.  Since it is a Markov process and since $Q$ is
distributed as a Gaussian asymptotically (up to corrections which we will
discuss), we will model the evolution of topological charge as an OU process.

The friction parameter $\eta$ is sensitive to algorithmic details and
therefore not physically relevant, so it will be treated as a nuisance
parameter.  Although we will not investigate it in detail here, we note that the parameter $\eta$
may be of interest in the comparison of different lattice update algorithms (with the underlying
physical parameters held fixed).  In particular, the autocorrelation $R(\tau)$ for the process $x(t)$ from the standard solution is given by
\beq
R(\tau) = e^{-\eta \tau} \left( \frac{1 - e^{-2\eta (t+\tau)}}{\sqrt{(1-e^{-2\eta t})(1-e^{-2\eta (t+\tau)})}}\right)
\eeq
which for $t \gg 1/\eta$ converges to $e^{-\eta \tau}$.  We can therefore identify $1/\eta$ as the standard autocorrelation time for $x$.  Our approach to be described below therefore gives an alternate way to estimate the autocorrelation time for an observable which is expected to always be approximately Gaussian distributed.

%%%%%%%%%%%%%%%%%%%%%%%%%%%%%%%%%%%%%%%%%%%%%%%%%%%%%%%%%%%%%%%%%%%%%%%%%%%%%%
\section{Maximum Likelihood Estimate \label{sec:ML}}
%%%%%%%%%%%%%%%%%%%%%%%%%%%%%%%%%%%%%%%%%%%%%%%%%%%%%%%%%%%%%%%%%%%%%%%%%%%%%%

We assume that we have $N+1$ computations of the topological charge $Q_i$ at
steps $n_i$ in the Markov chain, where the $n_i$ need not be equally spaced.
Due to parity invariance of Yang-Mills theory, all odd moments are identically
zero, including the mean $\langle Q \rangle = 0$.  The second moment, or
equivalently the variance, gives the topological susceptibility
\beq
\langle Q^2 \rangle = V \chi_t.
\eeq
In the OU model, we identify the susceptibility in terms
of the model parameters, $V \chi_t = \sigma^2 / 2 \eta$.  The conditional
probability of finding $Q_i$ at step $n_i$ given $Q_{i-1}$ was found at step
$n_{i-1}$ is Gaussian with mean and variance given by Eq.~(\ref{eq:OUvarmean})
with appropriate asymptotic values:
\begin{eqnarray}
\lefteqn{P(Q_i|Q_{i-1})} \nonumber \\*
& = & (2 \pi)^{-\frac{1}{2}}
\left[ V \chi_t \left( 1 - e^{-2 \eta (n_i-n_{i-1})} \right)
\right]^{-\frac{1}{2}} \nonumber \\*
&& \times \exp\left\{ - \frac{
  \left[ Q_i - Q_{i-1} e^{-\eta(n_i - n_{i-1})} \right]^2
}{
  2 V \chi_t \left[ 1 - e^{-2 \eta  (n_i-n_{i-1})} \right]
}
\right\} \ .
\end{eqnarray}
The log-likelihood function (dropping additive constants) given the time
series is
\begin{eqnarray}
L(\eta,V\chi_t) & = & - \frac{N}{2} \log V \chi_t
- \frac{1}{2 V \chi_t} S(\eta)
\nonumber \\*
&& - \frac{1}{2} \sum_{i=1}^N \log\left[ 1 - e^{-2 \eta (n_i - n_{i-1})}
\right]
\end{eqnarray}
where for later convenience we have defined the sum
\begin{equation}
S(\eta) \equiv \sum_{i=1}^N \frac{
  \left[ Q_i - Q_{i-1} e^{-\eta(n_i - n_{i-1})} \right]^2
}{
  1 - e^{-2 \eta  (n_i-n_{i-1})}
} \ .
\end{equation}
The maximum likelihood (ML) estimates $\widehat{\eta}$ and $\widehat{\chi}_t$
minimize the log-likelihood function $L(\eta, V \chi_t)$.  At the minimum,
\begin{equation}
\left .\frac{\partial L(\eta,V\chi_t)}{\partial V \chi_t}
\right|_{V \widehat{\chi}_t} = 0
\end{equation}
which leads to
\begin{equation}
V \widehat{\chi}_t = \frac{1}{N} S(\widehat\eta) \ .
\end{equation}
If we substitute $S(\eta)/N$ for $V \chi_t$ in the log-likelihood function we
now need to solve the one-dimensional problem to find $\widehat\eta$ that
minimizes
\begin{equation}
\mathcal{L}(\eta) = - \frac{N}{2} \log S(\eta)
- \frac{1}{2} \sum_{i=1}^N \log\left[ 1 - e^{-2 \eta (n_i - n_{i-1})} \right]
\end{equation}
where we have dropped further additive constants.  Once $\widehat\eta$ is
known then $V \widehat{\chi}_t$ is known as well.

\begin{figure}[t]
\includegraphics[width=0.45\textwidth]{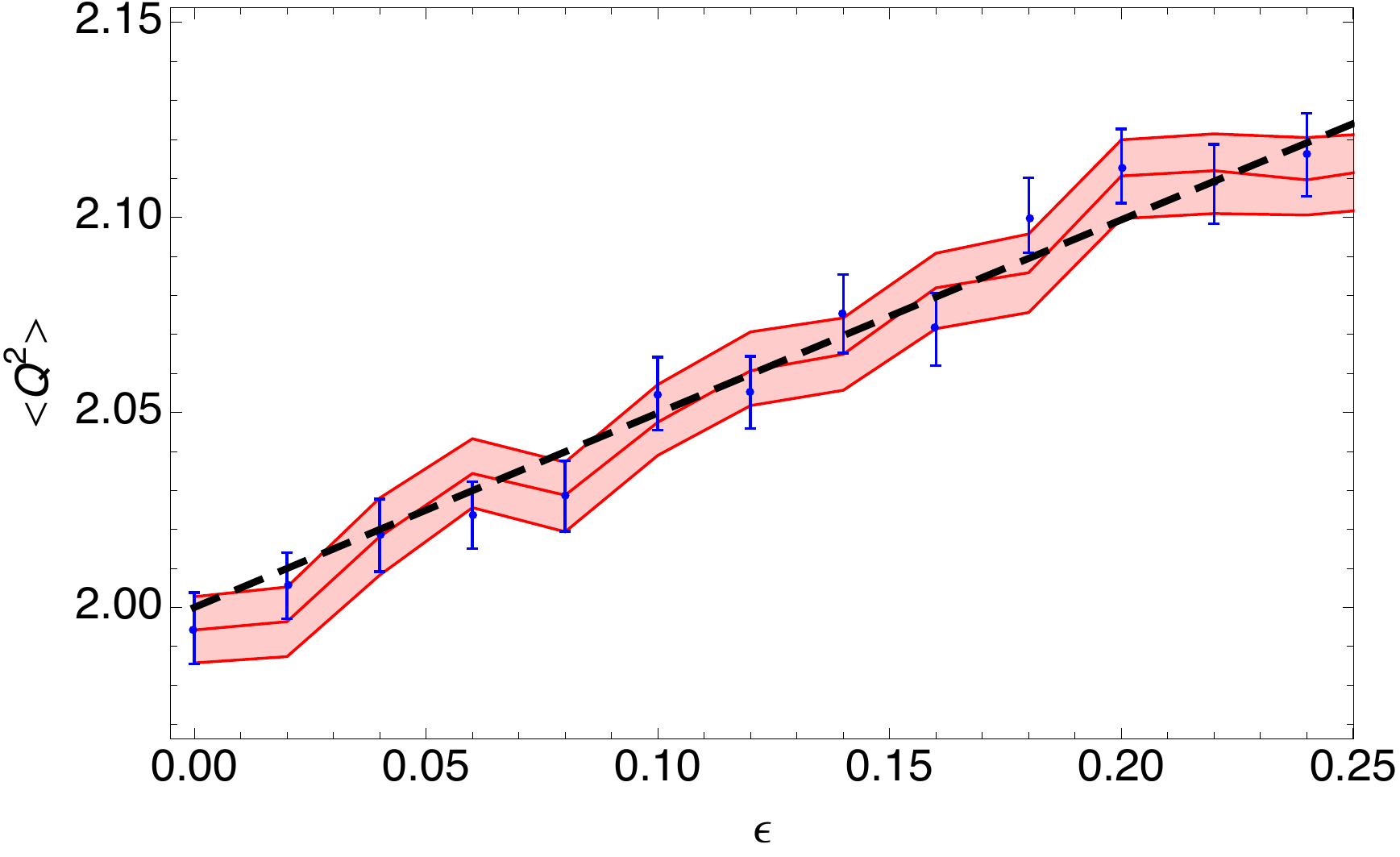}
\caption{\label{fig:edgeworth_variance_test}Monte Carlo test of variance extracted from an Edgeworth distribution \cref{eq:EdgeworthQ}, as a function of non-Gaussianity parameter $\epsilon$.  The dashed line (black) shows the analytic variance vs.\ $\epsilon$.  The points with error bars (blue) and the shaded band (red) show estimated variance using the sample variance and OU maximum-likelihood estimate, respectively.  Both methods show good agreement with the expected variance.}
\end{figure}

Since the OU model assumes the underlying distribution is Gaussian, it is
interesting to understand how well the OU-model ML estimates can reproduce
the variance of the nearly Gaussian distribution in Eq.~(\ref{eq:EdgeworthQ})
for $\epsilon \approx 0.2$ \cite{Giusti:2007tu}.  As a simple test, we generated 100,000 samples of
the distribution for $\sigma^2=2$ and $\epsilon = 0 \text{--} 0.24$ and
used both the OU ML method and the standard sample variance to estimate $\langle Q^2
\rangle$.  Both estimates agree well with the analytic value, as shown in \cref{fig:edgeworth_variance_test}.  In addition, near-perfect agreement is seen between the OU model and the standard sample variance, for this test in which the underlying true distribution is near-Gaussian and well-sampled.

%%%%%%%%%%%%%%%%%%%%%%%%%%%%%%%%%%%%%%%%%%%%%%%%%%%%%%%%%%%%%%%%%%%%%%%%%%%%%%
\section{NEARLY FIXED TOPOLOGY \label{sec:fixedtop}}
%%%%%%%%%%%%%%%%%%%%%%%%%%%%%%%%%%%%%%%%%%%%%%%%%%%%%%%%%%%%%%%%%%%%%%%%%%%%%%

%It has long been understood in lattice gauge theory calculations that the
%probability of a change in topological charge during a typical Markov step
%will decrease as a very high power of the lattice spacing $a$, see
%\cite{Luscher:2010iy} for a recent demonstration.  It is a reasonable concern
%if a Markov process used to perform some lattice calculation cannot
%reasonably sample the topological charge distribution in a finite running
%time due to limited number of tunneling events that the resulting calculation
%will be biased.

%In \cite{Brower:2003yx,Aoki:2007ka} it was suggested that one could still
%perform reliable calculations even if tunneling never occurred and the global
%topological charge remained fixed.  This method still requires a
%determination of $V \chi_t$ and it will clearly be difficult to determine it
%from a LS method if the number of tunneling events is $O(10)$ or less.  The
%ML method can certainly be used to make an estimate of $V \chi_t$ but given
%the discussion of Fig.~\ref{fig:plot-rbc01-ts}, this should probably be
%viewed as a lower bound.  Nonetheless, such a lower bound could prove useful
%in relating results computed in different topological sectors.

For lattice calculations in which the topological charge tunnels frequently,
the distribution of $Q$ will be well-sampled, and $\chi_t$ can be estimated
simply from the empirical sample variance, or from a least-squares (LS) fit of a
Gaussian to the $Q$ distribution.  The advantage of the ML method is that it
should still yield robust estimates of $\chi_t$ even when the distribution is
relatively poorly sampled.  However, in extreme cases where the number of
observed tunneling events is $O(10)$ or less, the uncertainty in $\chi_t$ can
become very large, as a lack of tunneling events can be explained by either
large $\chi_t$ and small $\eta$, or vice-versa.  Marginalizing over $\eta$
leads to essentially a lower bound on $\chi_t$.

Recently, it has been suggested that the use of Neumann boundary conditions
along one of the directions of the lattice would eliminate the barrier to
changing topology \cite{Luscher:2011kk}.  One can think rather informally of
this scenario as topological charge being allowed to flow freely through the
boundaries between the lattice and an infinite reservoir.  This physical
picture suggests an alternative approach to estimation of $\chi_t$.

Consider a periodic lattice with volume $V=L^3 \times T$ and $T \gg L$, and
then select some contiguous interval of length $T_s \ll T$, such that $V \gg
V_s \equiv L^3 \times T_s \gg \chi_t^{-1}$.  The total topological charge
$Q_s$ contained within this subvolume $V_s$ will be a continuous variable,
since charge is no longer conserved; it can move freely into the complement of
$V_s$, which we can think of as a reservoir.

The existence of a non-zero global topological charge $Q$ on the full volume
may bias the distribution of charge within the subvolume; in particular, if
$Q$ is fixed, then the mean charge contained within $V_s$ will be equal to $Q
V_s / V$.  We therefore define a ``subtracted'' subvolume charge,
\begin{equation}\label{eq:Qsub}
\Qssub = \int_{x \in V_s} d^4x \left[ q(x) - \frac{Q}{V} \right]
\end{equation}
where $q(x)$ is the topological charge density at lattice site $x$.  We then
carry out the analysis exactly as before, but with the substitutions $V \rightarrow V_s$ and $Q \rightarrow Q_{s,\textrm{sub}}$.

%\draftnote{EN: add argument about connection between $\chi_t$ and connected correlation function of $Q$?}  \draftnote{EN: no difference with subtraction found on RBC lattices...}
%
%\begin{equation}
%P(\Qssub) = \frac{1}{\sqrt{2 \pi V_s \chi_t}}
%\exp\left(-\frac{\Qssub^2}{2 V_s \chi_t} \right)
%\end{equation}
%

It seems reasonable, although not proven, that $\chi_t$ computed this way
is an acceptable estimator of topological susceptibility when using the
methods suggested in \cite{Brower:2003yx}, given that $V$ was periodic and
translationally invariant and $V_s$ was chosen at random.  Thus, we can apply
our same ML method to a time series in $Q_s$ to get an estimate of $V_s
\chi_t$.  Even with nearly-fixed $Q$, it may be possible for $Q_s$ to fluctuate
frequently enough to allow a reliable LS fit.  In this case, we can check that
ML and sample variance methods produce compatible results for $V_s \chi_t$.

%\draftnote{EN: out of order/not needed}
%\draftnote{DS: agreed and commented out}
%As a test of this idea we return to the ensembles shown in
%Fig.~\ref{fig:topology}.  Since the gauge configurations are freely available,
%we downloaded them and recomputed the topological charge density which we are
%free then to integrate over some subvolume $V_s$.

%%%%%%%%%%%%%%%%%%%%%%%%%%%%%%%%%%%%%%%%%%%%%%%%%%%%%%%%%%%%%%%%%%%%%%%%%%%%%%
\section{EXAMPLES \label{sec:examples}}
%%%%%%%%%%%%%%%%%%%%%%%%%%%%%%%%%%%%%%%%%%%%%%%%%%%%%%%%%%%%%%%%%%%%%%%%%%%%%%

\begin{figure}[t]
\includegraphics[width=0.45\textwidth]{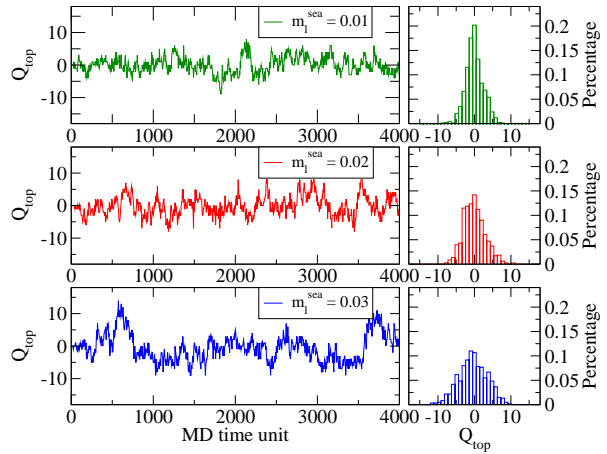}
\caption{\label{fig:topology}From
\cite{Allton:2007hx}, Markov-chain Monte Carlo time-series (left) and cumulative distributions (right) of global topological charge $Q$, as measured by the RBC and UKQCD collaborations.  Three ensembles are shown, differing by the light-quark mass: $m_l = 0.01$ (top), 0.02 (middle), and 0.03 (bottom).}
\end{figure}

As a trial of this method, we take a few time series of topological charge on a set of three
$16^3 \times 32$ lattice ensembles with $N_f = 2$+1 domain wall fermions, generated by the RBC and UKQCD collaborations \cite{Allton:2007hx}. The relevant
data and empirical distributions of $Q$ are plotted in \cref{fig:topology}.  For the analysis to follow we take a thermalization cut of 200 MD time units on all three ensembles.  The topological charge is measured every 5 MD time units.

\begin{figure}[t]
\includegraphics[width=0.45\textwidth]{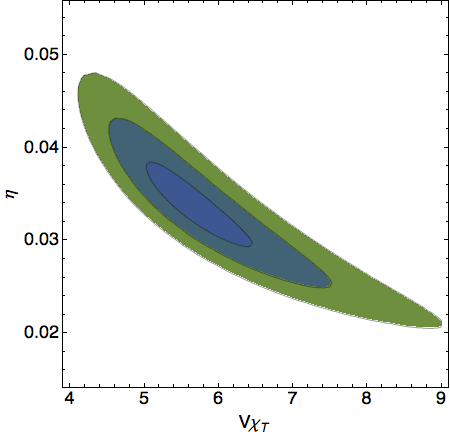}
\caption{\label{fig:rbc01-contour}Confidence contours for maximum likelihood analysis of the RBC/UKQCD $m_l=0.01$ ensemble, shown at $1\sigma$, $2\sigma$, and $3\sigma$ levels.}
\end{figure}

In Fig.~\ref{fig:rbc01-contour}, for the lightest mass $m_l = 0.01$
ensemble we show the $2 \Delta L = 1, 4, 9$ contours appropriate for
estimating the 1, 2, 3 $\sigma$ confidence intervals on $V\chi_t$ while
marginalizing over the friction parameter $\eta$.  The resulting 1-$\sigma$ confidence interval on $V\chi_t$ is found to be in good agreement with the standard sample-variance estimate.

The negative correlation between $V\chi_t$ and $\eta$ is expected, since they are inversely related through the asymptotic variance of the model distribution, $V\chi_t = \sigma^2 / 2\eta$. For data sets with relatively few tunneling events, we expect the ellipsoid will become elongated and follow a hyperbolic curve due to this relation.

We would now like to test the proposed subvolume analysis of \cref{sec:fixedtop}, in conjunction with both the sample variance and ML methods.  The use of only a fixed subvolume from all configurations would reduce the available statistics, so we make use of a bootstrap procedure in order to improve our statistical precision.  We draw $N_b = 1000$ bootstrap replications from the distribution of pairs $\{Q_i, Q_{i+1}\}$ in the topological charge time series, allowing us to resample while preserving the information on transitions required by the ML method.  We fix the subvolume size $T_s \leq T$, and then within each bootstrap replication choose a random starting position $t \in [0, N_t-1]$ for the subvolume on each configuration in the timeseries; the choice is randomized for each bootstrap replication.  This procedure imposes the expected translation invariance in the $t$-direction.

%First fixing the temporal extent $T_s \leq T$, we draw $N_b$ bootstrap replications of the $Q$ time series for calculation of the sample variance.  In the case of the ML method, the likelihood depends only on the transitions within $Q$, which we preserve by instead drawing from the distribution of pairs separated by one unit in the time series, $\{Q_i, Q_{i+1}\}$.  For the calculation of a single bootstrap replication of the likelihood, we draw a number of random positions $t$ equal to the length of the time series of $Q$ values (minus one if we are using pairs of $Q$ values), and take the $L^3\times T_s$ subvolume to start at that $t$ on each configuration in the time series.  This step imposes the expected translation invariance in the $t$-direction.

For the sample variance procedure, the data are blocked before drawing bootstrap samples, in order to deal with autocorrelation effects.  Empirical tests on the data show stability of error estimates on $\chi_t$ for a block length of roughly $\sim 100$ trajectories or 20 configurations.  No blocking is used for the ML analysis, which includes autocorrelation effects in the model.  For both the sample variance and ML subvolume analyses, the central value and error estimates correspond to the median and one-sigma quantiles of the bootstrap distribution, respectively.

%Subvolume variance: again, random subvolumes of full spatial extent and temporal extent $N_{t,s}$ are taken on each configuration.  The data are then blocked in order to deal with autocorrelation effects in the estimation of error bars; empirical tests on the data show stability of error estimates on $\chi_t$ for a block length of $\sim 100$ trajectories.  Blocks of length 100 trajectories (20 configurations) are used, except when considering the analysis with maximum number of configurations less than 600, in which case blocks of length 25 trajectories are used, in order to ensure an adequate number of data points are used in the sample variance to give negligible sample $1/N$ error on the variance.  2500 bootstrap replications are used to ensure stability in the $V \chi_t$ estimate due to $1/N_{boot}$ corrections. The same procedure is used for the full-volume bootstrap estimate, but without the subvolume selection at the first step. In both cases the central value and error estimates are given by median and one-sigma quantiles on the bootstrap distribution.

\begin{figure}[t]
\includegraphics[width=0.45\textwidth]{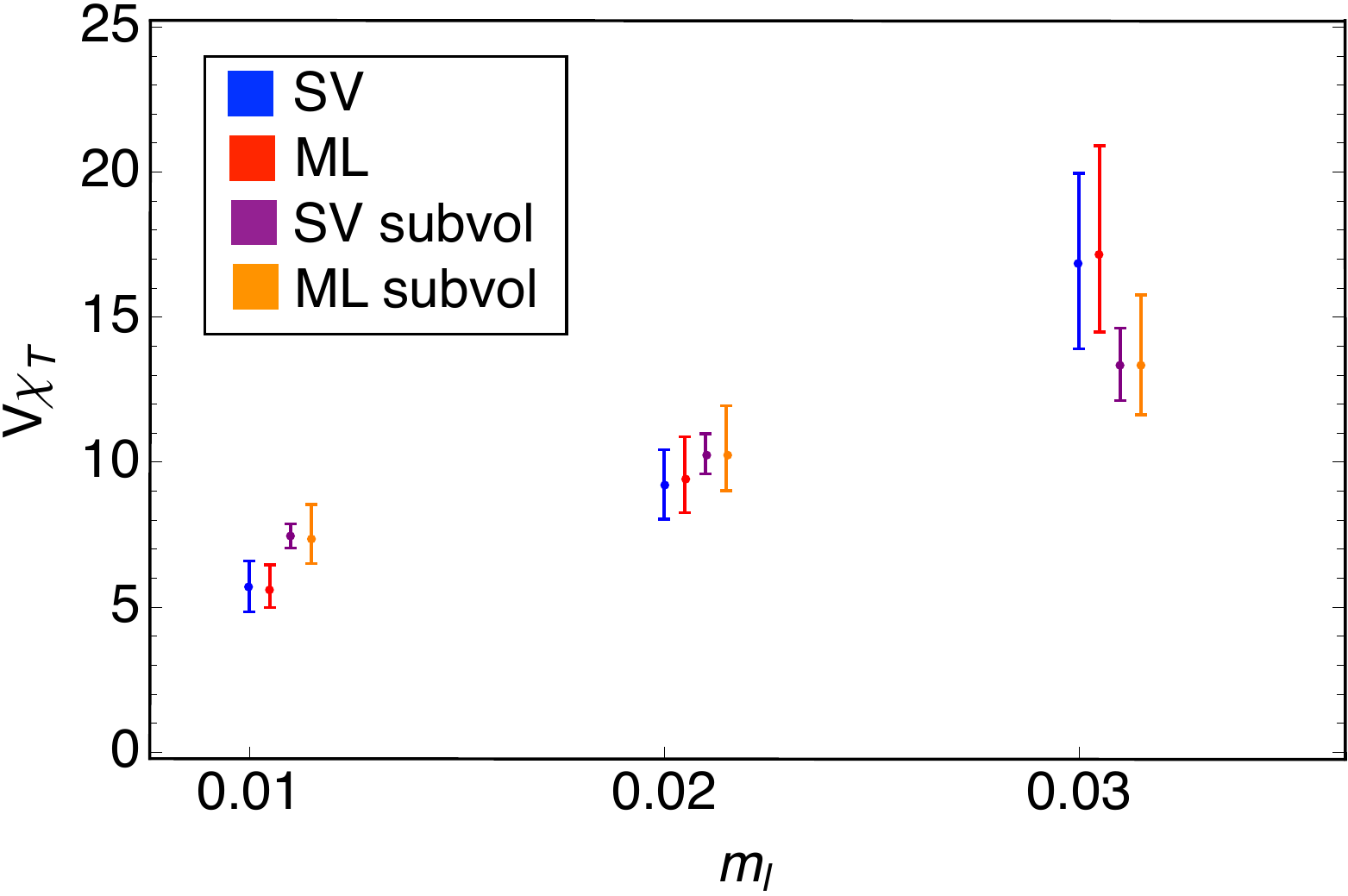}
\caption{\label{fig:rbc-compare} Comparison of various methods for determination of $V \chi_t$ on the three RBC/UKQCD example ensembles studied.  ``SV'' denotes use of the sample variance of $Q$, while ``ML'' indicates the maximum-likelihood method described in the text.  ``Subvol'' indicates analysis restricted to a subvolume with full spatial extent and $T_s = 8$.}
\end{figure}

\begin{table}
\begin{tabular}{|c|ccc|}
\hline
$V \chi_t$&$m_l = 0.01$&0.02&0.03\\
\hline
SV&5.68\errpm{+91}{-84}&9.19\errpm{+1.23}{-1.16}&16.86\errpm{+3.10}{-2.96}\\
ML&5.64\errpm{+82}{-66}&9.40\errpm{+1.47}{-1.14}&17.18\errpm{+3.73}{-2.70}\\
SV subvol&7.44\errpm{+43}{-41}&10.26\errpm{+71}{-66}&13.32\errpm{+1.30}{-1.20}\\
ML subvol&7.38\errpm{+1.16}{-0.87}&10.24\errpm{+1.71}{-1.22}&13.38\errpm{+2.38}{-1.75}\\
\hline
\end{tabular}
\caption{\label{tab:compare} Comparison of various methods for determination of $V \chi_t$ on the three RBC/UKQCD example ensembles studied.  ``SV'' denotes use of the sample variance of $Q$, while ``ML'' indicates the maximum-likelihood method described in the text.  Rows labeled ``subvol'' apply the same techniques on a subvolume with full spatial extent and $T_s = 8$.}
\end{table}

In Fig.~\ref{fig:rbc-compare} and \cref{tab:compare}, we summarize our determination of $V
\chi_t$ for the three ensembles shown in Fig.~\ref{fig:topology} using the various methods described.  The subvolume results here are for fixed $T_s = 8$.  As expected from a time series with many independent samples of $P(Q)$, the maximum likelihood (ML) result agrees closely with the sample variance (SV) estimate of $V\chi_t$.

\begin{figure}[h]
\includegraphics[width=0.45\textwidth]{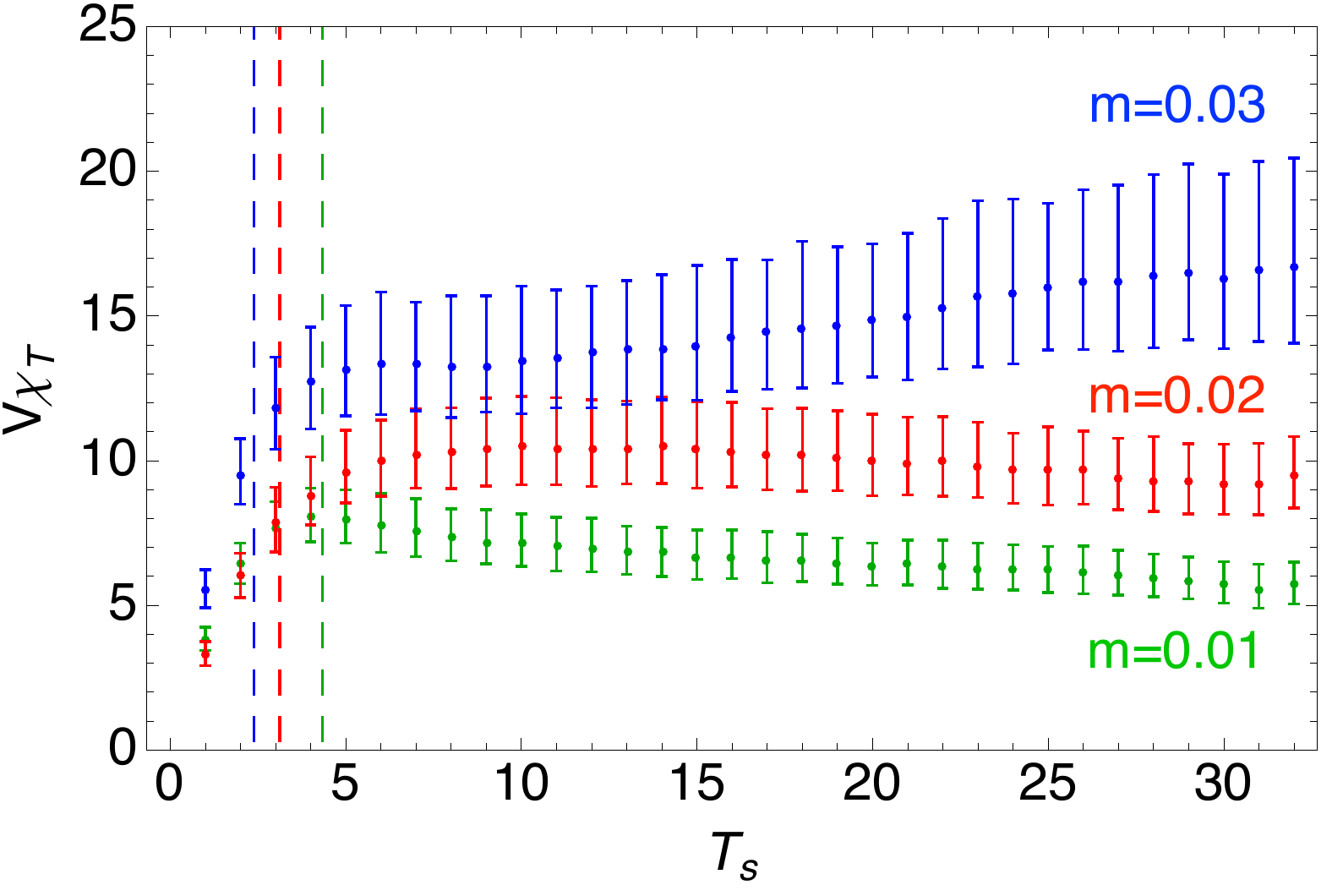}
\caption{\label{fig:subvol}Dependence of ML subvolume estimates of $V\chi_t$ on the temporal extent of the $L^3 \times T_s$ subvolume used, for the three RBC/UKQCD example ensembles, from bottom to top $m_l = 0.01$ (green), $0.02$ (red) and $0.03$ (blue).  The vertical lines show the point at which $V_s \chi_t \approx 1$ on each ensemble (based on the ML estimate at $T_s = 8$), beyond which our method is expected to break down.}
\end{figure}

We further investigate the subvolume estimates, and in particular their dependence on the choice of subvolume size, by varying the temporal extent $T_s$ and repeating the analysis.  The results are shown for all three ensembles in \cref{fig:subvol}.  A strong variation is seen at small $T_s$, setting in approximately where $V_s \chi_t \approx 1$, which is where our assumptions about the simplicity of the distribution $P(Q)$ are anticipated to break down.  For large $T_s$ the dependence on subvolume size is nearly flat, but with a small systematic trend evident, particularly on the $m_l = 0.03$ ensemble.  We have no immediate physical explanation for the origin of this subleading effect, but plan to investigate further in a future work.

\begin{figure}[h]
\includegraphics[width=0.45\textwidth]{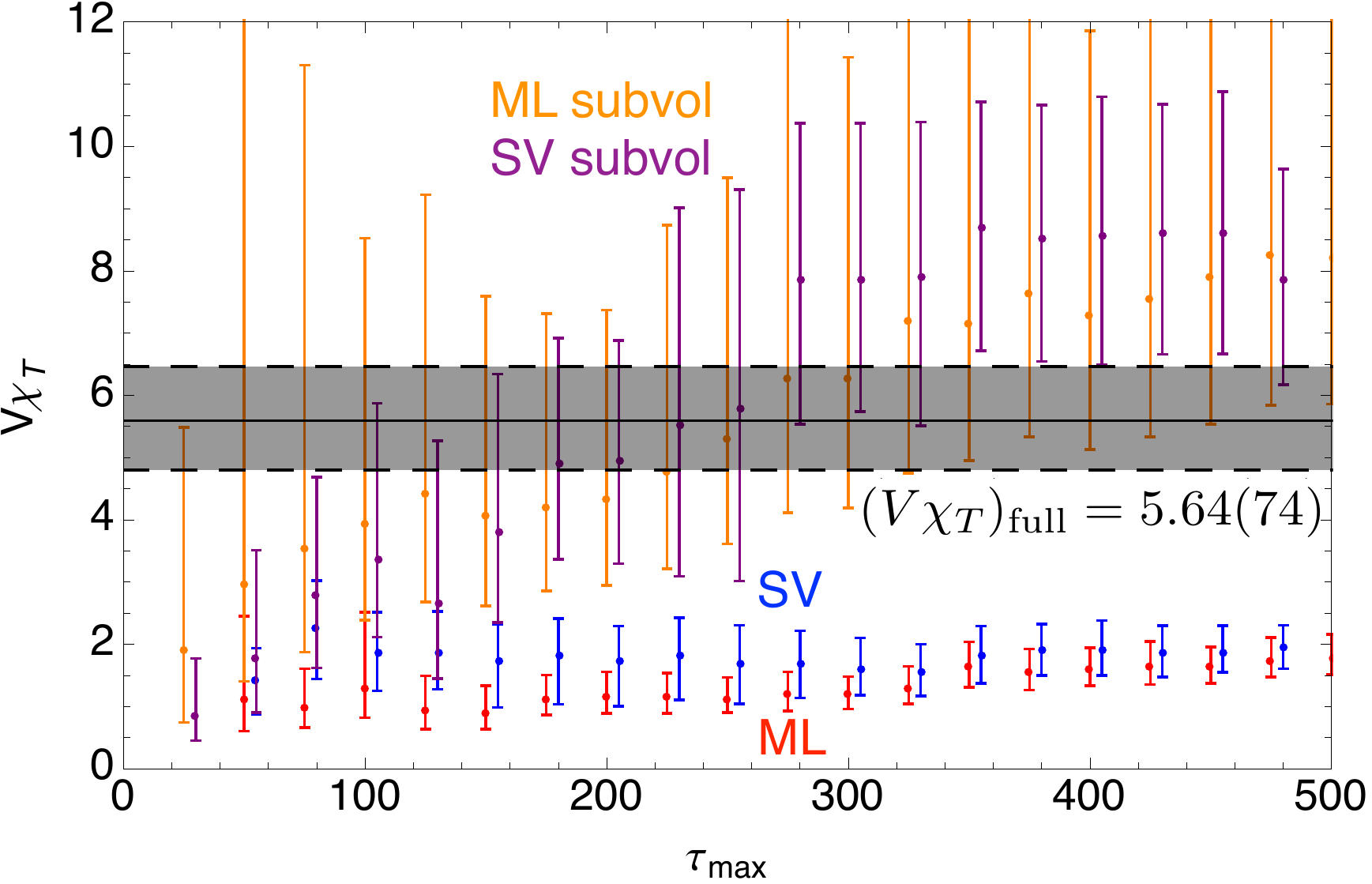}
\caption{\label{fig:plot-ts-compare}Comparison of estimates of topological susceptibility times the volume $V\chi_t$ taken on a subset of the RBC $m_l=0.01$ gauge configurations, with MD trajectory numbers up to $\tau_{\textrm{max}}$.  Points shown correspond to different methods: sample variance (blue), maximum likelihood (red), subvolume SV (purple), and subvolume ML (orange).  All SV-method results are shown with an offset for clarity.  Subvolume ML estimates (again with $T_s = 8$) show the best consistency with the ``asymptotic" result (black band) obtained by the ML method applied to the full time series.}
\end{figure}

It is apparent that the ML method does not offer any significant advantage in the determination of $V\chi_t$ over a simple calculation of the sample variance when the underlying distribution $P(Q)$ is well-sampled, as is the case for the full time series on each of the RBC ensembles.  However, we expect the ML technique to be a robust approach even when a small number of independent samples are available.  Furthermore, even when the distribution is well-sampled, the ML method has the advantage of including autocorrelation effects automatically, into the friction parameter $\eta$, whereas the SV analysis requires an autocorrelation analysis and blocking to be carried out first.

We can test what might happen in a case with poor sampling by analyzing a restricted subset of the RBC time series.  Fig.~\ref{fig:plot-ts-compare} shows the results of this test on the $m_l = 0.01$ ensemble, with the analysis considered on the restricted time series with MD time $\tau \leq \tau_{\textrm{max}}$; as a reminder, $Q$ is measured every 5 MD time units.  For the SV analysis, we adjust the blocking when only a small number of configurations are available; specifically, we use a block length of $\tau = 50$ when less than $200$ time units are available, and $\tau = 25$ for less than $100$ time units available.  With only a subset of the configurations, the full-volume methods show a clear bias with respect to the best estimate of $V \chi_t$ from the full ensemble.  On the other hand, both the ML and SV subvolume approaches converge rapidly to cover the asymptotic estimate, with the ML being particularly effective at small $\tau_{\textrm{max}}$ where a simple blocking analysis cannot adequately account for the known autocorrelation effects.

%To emphasize this point, again for the lightest mass ensemble, in Fig.~\ref{fig:plot-rbc01-ts} we show the ML estimate of $V\chi_t$ for a subset of the data computed on lattices with trajectory numbers less than some maximum number.  Clearly the estimates from the early part of the time series underestimate the susceptibility of the full time series.  This merely reflects the relatively small fluctuations in the early part of the time series shown in Fig.~\ref{fig:topology}.  It's only after the relatively large fluctuations occurring around trajectory 1250 and again after trajectory 1800 that the ML estimates will essentially stabilize around the final answer. Nonetheless, the ML method seems quite effective at quickly establishing a lower bound on the susceptibility after roughly ten tunneling events.

%\begin{figure}[t]
%\includegraphics[width=0.45\textwidth]{Vchi_compare_RBC01_full.pdf}
%\includegraphics[width=0.45\textwidth]{subvolume-compare-RBC-m01.pdf}
%\caption{\label{fig:plot-subvol-compare} Comparison of subvolume ML estimates of $V \chi_T$ vs. subinterval width $T_s$, again for the RBC $m_l=0.01$ gauge configurations.  The blue circles make use of the first 700 gauge configurations, while the red triangles include only the first 100 gauge configurations.}
%\end{figure}

%%%%%%%%%%%%%%%%%%%%%%%%%%%%%%%%%%%%%%%%%%%%%%%%%%%%%%%%%%%%%%%%%%%%%%%%%%%%%%
\section{Discussion \label{sec:discussion}}
%%%%%%%%%%%%%%%%%%%%%%%%%%%%%%%%%%%%%%%%%%%%%%%%%%%%%%%%%%%%%%%%%%%%%%%%%%%%%%

We have introduced a new maximum-likelihood approach to estimation of the topological susceptibility $\chi_t$ in lattice calculations, based on maximum-likelihood analysis of the full time-series information.  This approach can give an advantage over more traditional methods such as calculation of the sample variance of $Q$, particularly in the case that autocorrelation times are long and relatively few independent samples are available, due to the inclusion of autocorrelation effects within the ML model.  The autocorrelation time of $Q$ can also be estimated as a byproduct of the analysis.

In addition, we have explored the analysis of topological charge fluctuations on lattice subvolumes.  This technique may be necessary in cases where the global topological charge goes through few or even no tunneling events within a lattice calculation.  Even when $Q$ fluctuates adequately, the subvolume method (in conjunction with the ML analysis) was found to give the most robust estimates of $\chi_t$, with confidence intervals rapidly converging to cover the asymptotic estimates of this quantity even on small amounts of data.  Stability of the estimate with respect to the subvolume size was observed empirically down to $V_s \chi_t \approx 1$, at which point our physical assumptions about the fluctuations should break down.

A modification of the ML approach to a more complex model than the OU process, which might be able to deal with non-Gaussian distributions and therefore extract higher moments by the maximum-likelihood approach, would be interesting to study in a future work.  The modeling of higher-order systematic dependence on the subvolume size, as hinted at by our current analysis, also merits further study.

%\paragraph*{Acknowledgments}

We thank the RBC-UKQCD collaboration for the use of their lattice configurations, first published in Ref.~\cite{Allton:2007hx}, and we also thank Tom Blum and Philippe de Forcrand for useful discussions.  This work was supported in part by the National Science Foundation under Grant No.~PHYS-1066293 and the hospitality of the Aspen Center for Physics.  M.L.\ was partially supported by SciDAC-3 and Argonne Leadership Computing Facility at Argonne National Laboratory under contract DE-AC02-06CH11357, and the Brookhaven National Laboratory Program Development under grant PD13-003.  D.S.\ was supported by DOE Grant Nos.~DE-SC0010005, DE-SC0008669 and DE-SC0009998.  R.~C.~B., C.~R., and E.~W.\ were supported by DOE grant DE-SC0010025.  In addition, R.~C.~B., C.~R., M.~C. and O.~W.\ acknowledge the support of NSF grant OCI-0749300, and G.~F. and G.~V. were supported by NSF grant PHY11-00905.  We thank LLNL for funding from LDRD13-ERD-023, and E.~R., C.~S., and P.~V.\ acknowledge the support of the U.~S.~Department of Energy under Contract DE-AC52-07NA27344 (LLNL).

%%%%%%%%%%%%%%%%%%%%%%%%%%%%%%%%%%%%%%%%%%%%%%%%%%%%%%%%%%%%%%%%%%%%%%%%%%%%%%
\bibliography{OUtop}
%%%%%%%%%%%%%%%%%%%%%%%%%%%%%%%%%%%%%%%%%%%%%%%%%%%%%%%%%%%%%%%%%%%%%%%%%%%%%%

\end{document}